\documentstyle[12pt]{article}
\begin{document}

\baselineskip 0.9cm

\begin {center}{\LARGE {Gravitational and Particle Radiation
        from Cosmic Strings }}
\end{center}
\vspace{1.cm}
\centerline {\it Anzhong Wang \footnote{e-mail address: wang@on.br} and
Nilton O. Santos \footnote{e-mail address: nos@on.br}}

\begin{center}
Department of Astrophysics, Observat\'orio Nacional--CNPq, Rua General
Jos\'e Cristino 77, S\~ao Crist\'ov\~ao, 20921-400 Rio de Janeiro, Brazil
\end{center}

\vspace*{1cm}

\begin{abstract}

\baselineskip 0.9cm

Gravitational and massless particle radiation of straight cosmic
strings with finite thickness is studied analytically. It is found that
the non-linear interaction of the radiation fields emitted by a cosmic string
with the ones of the string always makes the spacetime singular
at the symmetry axis. The singularity is not removable and is a scalar
one.
\vspace{1.5cm}


\end{abstract}

\newpage

\baselineskip 0.9cm

Gauge theories predict the existence of topological defects, such as,
domain walls, cosmic strings, and monopoles, in the early stages of the
Universe [1]. Among them, the most studied are cosmic strings (CS), because
they may provide seeds for the formation of galaxies and the large
scale structure of the Universe [2]. This scenario of the structure
formation has been extensively studied in the past decade [3], and some
of its fundamental assumptions have been challenged by some recent
numerical simulations [4]. In particular, it was found that there is
a significant amount of small-scale structure on scales from the
horizon down to the limit of resolution of the simulations. As a
result, the single-scaling regime, upon which the original scenario
rests, seems to be ruled out. Most of the small-scale structure is due
to kinks, which are formed on the long CS when they cross and
intercommute. These kinks emit gravitational radiation, which is very
similar, in both frequency and amplitude, to the one from loops [5].

The study of the gravitational radiation of CS is of
particular importance to the  cosmological scenario, since most of the
observational
limits on CS networks in the early Universe are obtained
from it, and more important, the  multiple-scaling regime [3] crucially
depends on the rate of the radiation. If the radiation is strong enough
so that the small-scale structure can be smoothed out by it, a
multiple-scaling is resulted, and the scenario can be saved.
Thus, to have the scenario work, a complete understanding of the
CS gravitational radiation is required.

In this Letter we shall study the backreaction of the gravitational and
particle radiation emitted by infinitely long cosmic strings (ILCS), by
considering analytical models to the Einstein field equations. In order
to avoid some mathematical uncertainties [6], we shall consider ILCS
with finite thickness. In addition, our study of particle radiation is
only restricted to the massless one.  The main reason is that for
macroscopically large CS, only can the emission of massless
particles be important, and the ones of particles with masses are
usually suppressed [1].

It should be noted that the gravitational interaction of ILCS with
parallel  moving gravitational and electromagnetic waves has been
studied in [7, 8]. In particular, Letelier [8] showed that the
interaction sometimes could produce intermediate (or non-scalar)
singularities. However, since there is no momentum transported away
from the CS, his model as well as the one studied in [7] cannot
be considered as radiation of CS. The
gravitational radiation of ILCS was studied in [9] by weak-field
approximation (WFA). Again, because of the use of the WFA,
the backreaction of the radiation field was not fully taken into account.
Due to the non-linearity of the Einstein field equations, however, it
is expected that it might play a significant role in the evolution of CS.
In fact, in this Letter we shall show that it is this backreaction
that {\em always} makes the spacetime of CS singular on the
symmetry axis.

To begin with, let us first consider the spacetime with the metric [10]
\begin{equation}
ds^2 = e^{-b(\eta)} (dt^2 - dr^2) - dz^2 - r^{2} d\varphi^{2} ,
\end{equation}
where $t, r, z,$ and $\varphi$, as cylindrical coordinates, have their
usual physical meaning, with $- \infty < t, z < + \infty, 0 \le r < +
\infty,$ and $ 0 \le \varphi \le 2\pi$. And $b(\eta)$ is given by
$b(\eta) = [B(\eta) - B(\eta_{0})]H(\eta- \eta_{0}),$ where $\eta
\;(\equiv t-r)$ denotes the retarded time, $\eta_{0}$ is an arbitrary
constant, and $H(x)$ the Heaviside function which is $1$ for $x > 0$
and $0$ for $x < 0$. $B(\eta)$ is a smooth and monotonically
increasing function [Such a requirement comes from the consideration of
the energy-momentum
tensor (EMT). See the discussions given below.]. Clearly,
when $B(\eta) = 0$, the metric represents
the Minkowski spacetime written in a cylindrical coordinate system.
When $B(\eta) \ne 0$, it represents an exploding cylindrical null dust
cloud accompanied by a cylindrical gravitational wave. To see this, let
us first consider the corresponding  EMT, which
is now given by [11] $T_{\mu\nu} = B'(\eta)H(\eta - \eta_{0})/(2r)
k_{\mu} k_{\nu}$, where $k_{\mu}$ is a null vector defined as $k_{\mu}
\equiv \delta^{t}_{\mu} - \delta^{r}_{\mu}$. A prime denotes the
ordinary differentiation with respect to the indicated argument. It is
clear that this EMT represents a null dust cloud, consisting of
massless particles and moving along the null hypersurfaces $\eta = const.$
The presence of a gravitational wave can be seen by calculating the
Weyl tensor, $C_{\mu\nu\lambda\sigma}$, which is thought of as
representing the purely gravitational field. In the present case, it
turns out that the only non-vanishing component of it is $\Psi_{0}
\equiv - C_{\mu\nu\lambda\sigma} L^{\mu}M^{\nu}L^{\lambda}M^{\sigma} =
- e^{b(\eta)} B'(\eta)H(\eta - \eta_{0})/(2r)$, where $L^{\mu} \equiv
e^{b(\eta)/2}(\delta_{t}^{\mu}-\delta_{r}^{\mu})/\sqrt{2}$ and $M^{\mu}
\equiv ( \delta_{z}^{\mu} + ir^{-1} \delta_{\varphi}^{\mu})/\sqrt{2}$,
which means that the spacetime is Petrov type N, a typical
gravitational wave spacetime [12]. In the present case, the
gravitational cylindrical wave moves outward along the
surfaces $\eta = const.$  From the expressions of $T_{\mu\nu}$ and
$\Psi_{0}$, we can see that the outgoing gravitational wave and dust
cloud exist only after the time $t = \eta_{0} $, since $H(\eta-
\eta_{0})$ is zero for $t <\eta_{0}.$

On the other hand, it was shown in [13] that for a static
CS  the spacetime outside it is described by a
flat cylindrical metric with a wedge of the angular size
$\delta\varphi = 8\pi \mu$ being removed, where $\mu$ is the linear mass
density per unit length. Thus, the superposition of the above two
spacetimes will be the metric (1) with $d\varphi$ being  replaced by
$(1-4\mu)d\varphi$. Obviously, the resulting metric represents a
cylindrical null dust cloud plus a gravitational wave propagating outwards
in a CS background. Due to their presence
the angle defect now is
\begin{equation}
\delta \varphi = 2\pi \left[ 1 - (1 - 4\mu)
r/\int_{0}^{r}e^{-b(\eta)/2}dr\right],
\end{equation}
which is always less
than $8\pi \mu$. This happens because both, the dust
cloud and the gravitational wave, carry away positive energy. The latter
can be seen from Thorne's $C-energy$ [14], which here is given by ${\cal
C} = [1-(1-4\mu)^{2}e^{b(\eta)}]/8.$ Thus, the rate of the energy loss
is ${\cal C},_{\eta} = -
(1-4\mu)^{2}e^{b(\eta)}b'(\eta)/8$, which is
non-positive. That is, the outgoing gravitational wave and dust cloud
always carry away positive energy and  make the angle defect smaller.

Taking the above solution as valid only in the region $r \ge r_{0} > 0$,
where $r_{0}$ is the coordinate radius of the CS, and then
extending it into the CS, we finally obtain a solution which can be
written in the form of Eq.(1) but with the metric coefficient
$g_{\varphi\varphi}$ being replaced by
\begin{equation}
 g_{\varphi\varphi} = - w^{2} = -  \beta(r)^{2}H(r_{0} - r) -
(1 - 4\mu)^{2}r^{2}H(r - r_{0}),
\end{equation}
where $\beta(r)$ is a smooth function with the boundary conditions
$\beta(r_{0}) = (1 - 4\mu)r_{0},$ and $\beta'(r_{0}) = (1 - 4\mu)$,
which  assure that the hypersurface $r = r_{0}$ is a regular boundary
surface.  When $B(\eta) = 0$, the above solution reduces to the general
static CS solution studied by Linet in [13]. When $B(\eta)
\ne 0$, it represents a CS, which is static for $t <
\eta_{0}$ and starts to emit gravitational waves and massless particles
at the moment $t = \eta_{0}$. To show this explicitly, let us first
consider the corresponding EMT. It is easy to find
\begin{equation}
 T_{\mu\nu} = \sigma (g_{\mu\nu} + r_{\mu}r_{\nu} + \varphi_{\mu}\varphi_{\nu})
              + \rho k_{\mu}k_{\nu},
\end{equation}
$$\sigma = - e^{b(\eta)} \beta'' \beta^{-1} H(r_{0}-r),\;\;\;
\rho = B'H(\eta-\eta_{0})w,_{r}/(2w)$$
\begin{equation}
w,_{r}/w =  \beta'\beta^{-1} H(r_{0}-r) + r^{-1} H(r-r_{0}),
\end{equation}
where $ r_{\mu} = (-g_{rr})^{1/2}\delta^{r}_{\mu}, \varphi_{\mu} =
(-g_{\varphi\varphi})^{1/2} \delta^{\varphi}_{\mu}.$  Thus, providing
that $\beta''(r) < 0$, the first term in r.h.s. of Eq.(4) represents a
CS with support in the region $0 \le r \le r_{0}$. The
regularity conditions at the axis $r = 0$ require
\begin{equation}
\beta(r) \rightarrow r[ 1 - O(r^{2}) ],\;\;\;\;\;\;\;\;\;
as \;\; r \rightarrow 0.
\end{equation}
To have the circumference monotonically increasing as $r$
increases, we further require $\beta'(r) > 0$ (cf. Linet in [13]).  The
second term represents a null dust cloud moving outward from the axis.
Since $\rho$ is proportional to $H(\eta-\eta_{0})$, we can see that
this cloud comes to exist only after the moment $t = \eta_{0}$. Note
that the energy density of it is continuous across the hypersurface $r
= r_{0}$.  On the other hand, from the equations
$T_{\mu \lambda}\;^{;\lambda} = 0$ we find $T_{\mu
\lambda}^{cs;\lambda} = - (\rho k_{\mu}k_{\lambda})^{;\lambda}
=\frac{1}{2}\sigma b'(\eta)k_{\mu},$ where $T_{\mu \lambda}^{cs}$ is
the EMT for the CS given by the first part in r.h.s. of Eq.(4),
and $\sigma$
is given by Eq.(5). The above expression shows clearly that the
CS is indeed the source of the emission. Outside the CS,
$T_{\mu \lambda}^{cs;\lambda} = 0$, and the dust cloud propagates to
the space-like infinity with its amplitude decreasing like $r^{-1}$.
At a moment, say, $t$, the energy per unit length inside the radius $r$, is
\begin{equation}
{\cal C} = \left\{ 1 - [\beta'^{2}H(r_{0} - r) +
(1 - 4\mu)^{2}H(r - r_{0})]e^{b(\eta)}\right\}/8,
\end{equation}
from which we have that ${\cal C},_{\eta}$ is always non-positive.
That is, the CS is losing its energy through radiation. The
maximum amount of the energy loss is ${\cal C}_{max} = [1-(1 - 4\mu)^{2}]/8
\;(\approx \mu$ when $\mu \ll 1)$.  This gives a limit on the function
$b(\eta)\; \le -2ln(1-4\mu).$ When it is greater than this limit, the
outgoing wave and cloud will carry away energy more than the CS
has. Then, the CS will effectively have negative mass [cf. Eq.(7)].
In the following we shall not consider this possibility.

On the other hand, introducing a null tetrad frame by
$$ L_{\mu} = (g_{tt})^{1/2}(\delta^{t}_{\mu} +
\delta^{r}_{\mu})/\sqrt{2},
M_{\mu} = [(-g_{zz})^{1/2}\delta^{z}_{\mu} +
i(-g_{\varphi\varphi})^{1/2}\delta^{\varphi}_{\mu}]/\sqrt{2},$$
\begin{equation}
N_{\mu} = (g_{tt})^{1/2}k_{\mu}/\sqrt{2},
\bar{M}_{\mu} = [(-g_{zz})^{1/2}\delta^{z}_{\mu} -
i(-g_{\varphi\varphi})^{1/2}\delta^{\varphi}_{\mu}]/\sqrt{2},
\end{equation}
where $k_{\mu}$ is defined as in (4), we find that the non-vanishing
components of the Weyl tensor in this frame are
$$\Psi_{0} \equiv - C_{\mu\nu\lambda\sigma}
          L^{\mu}M^{\nu}L^{\lambda}M^{\sigma}
          = \Psi^{cs}_{0} + \Psi^{gv}_{0}, $$
$$\Psi_{4} \equiv  - C_{\mu\nu\lambda\sigma}
N^{\mu}\bar{M}^{\nu}N^{\lambda}\bar{M}^{\sigma}
          = \Psi^{cs}_{4} ,$$
\begin{equation}
\Psi_{2} \equiv  - C_{\mu\nu\lambda\sigma} L^{\mu}N^{\nu}
          (L^{\lambda}N^{\sigma}
          - M^{\lambda}\bar{M}^{\sigma})/2
          = \Psi^{cs}_{2},
\end{equation}
where
$$\Psi^{cs}_{0} = \Psi^{cs}_{4} = - 3 \Psi^{cs}_{2}
  =  e^{b(\eta)}\beta''/(4 \beta)H(r_{0}-r),$$
\begin{equation}
\Psi^{gv}_{0} =  -e^{b(\eta)}b'(\eta)w,_{r}/(2w).
\end{equation}
The reason for calculating the Weyl tensor in the null frame
(8) is that its components in this frame have direct physical
meaning, for example, $\Psi_{0}$ represents the gravitational wave
component propagating along the null hypersurfaces defined by $N_{\mu}$
( or equivalently $k_{\mu})$, and $\Psi_{4}$ the component propagating
along the null hypersurfaces defined by $L_{\mu}$ [12]. Equation (9)
shows that $\Psi_{0}$ contains two terms, the first is finite and
has support only
inside the CS, and so do the $\Psi_{2}$ and $\Psi_{4}$ terms. We
interpret them as the gravitational field of the CS. The second
term of $\Psi_{0}$ is proportional to $b'(\eta)$ and
has support in the whole spacetime. Moreover, it is
continuous across the hypersurface $r = r_{0}$. Outside the
CS, this term has a typical form of a gravitational wave. So, we
interpret it as representing the gravitational radiation of the CS.
Therefore, we conclude that the solution given by (1) and (3)
indeed represents the gravitational and massless particle emission of a
CS. Note that although the CS is emitting gravitational waves
and massless particles, Eq.(5) shows that the energy density of it
 still increases as the time develops [recall that
$b(\eta)$ is a monotonically increasing function]. This is because,
as the CS is radiating, it also undergoes a collapse.
The proper radius of it now is given by $\int_{0}^{r_{0}}
e^{-b(\eta)/2} dr$, which decreases as the time passes. However,
since $b(\eta)$ is a function with a limit, it never collapses into a
singular line.

{}From Eqs.(1) and (3) we can show [10] that the null
vector $k_{\mu}$ defines a null geodesic congruence, and that the
quantity $Q_{k} \equiv - k^{\lambda};_{\lambda} = -e^{b(\eta)}
w,_{r}/w,$ represents
the rate of the contraction of the  congruence.
Considering  Eqs. (5) and (6) we find that $Q_{k}\rightarrow - \infty$ as $r
\rightarrow 0$, which indicates that when the backreaction of the
radiation is taken into account, the axis might become singular. To
show that this is indeed the case, let us turn to consider the
Kretschmann scalar, which is found to take the form
\begin{equation}
{\cal R} \equiv R_{\mu\nu\lambda\sigma}R^{\mu\nu\lambda\sigma} =
8[8(\Psi^{cs}_{0})^{2} + \sigma^{2}]/3 + 16\Psi^{cs}_{4}\Psi^{gv}_{0}
+ 4e^{b(\eta)} \rho\sigma,
\end{equation}
where $\rho, \Psi^{cs}_{0}, \Psi_{4}$, and
$\Psi^{gv}_{0}$ are given, respectively, by Eqs.(5), (9) and (10).
Equation (11) contains three terms, each of which has the following
physical meaning: The first term represents the contribution of the
CS, which is finite. The second represents the interaction of
the gravitational field component $\Psi^{cs}_{4}$ of the CS with the
outgoing wave $\Psi^{gv}_{0}$, while the last represents the interaction
between the matter field of the CS and the outgoing null dust
cloud. From Eqs.(6), (9) and (10) we can see that the last two terms
become unbounded as $r\rightarrow 0$. That is, because of the
backreaction of both gravitational and massless particle radiation of
the CS, the spacetime becomes singular at the axis. This
singularity is a scalar one and always formed, as long as
the radiation is not zero, characterizing by the function $b(\eta)$. A
particular case is $\beta(r) = \alpha
\sin(\delta r)$, where $\alpha$ and $\delta$ are two constants and
determined by the boundary conditions at $r=r_{0}$. When $b(\eta) = 0,$
the solution reduces to the one studied by Hiscock in [13]. By properly
taking the limit $r_{0} \rightarrow 0$, Hiscock obtained the solution
of an infinitely thin CS with support only on the axis. Equation (10)
shows that after taking this limit the $\Psi^{cs}_{A}$'s are still
different from zero and have support on the axis, too. This observation
is important, since it indicates that our above results concerning the
singularity behavior also hold for the case of an infinitely thin
CS radiating gravitational waves, a case often considered in
the numerical stimulations of long CS networks [3].

On the other hand, let us consider the timelike geodesics perpendicular to
the CS. The first  integration of the geodesic equations yields
$dt/d\tau = (1+e^{b})/2, dr/d\tau = (1-e^{b})/2, dz/d\tau = 0,$ and
$d\varphi/d\tau = 0,$ where $\tau$ is
the proper time. Perpendicular to the tangential vector $\lambda^{\mu}_{(0)}$
$(\equiv
dx^{\mu}/d\tau),$ we can construct  other three linearly independent
spacelike vectors $\lambda^{\mu}_{(a)} ( a = 1, 2, 3)$ by
$\lambda^{\mu}_{(1)} =
(dr/d\tau)\delta^{\mu}_{t} + (dt/d\tau)\delta^{\mu}_{r},
\lambda^{\mu}_{(2)} = \delta^{\mu}_{z},$ and $\lambda^{\mu}_{(3)}
 = (1/w)\delta^{\mu}_{\varphi}$. One can show
that these four unit vectors form a freely-falling frame.
Computing the Riemann tensor in this frame, we find that it has only two
independent components, one of them being given by
$$
R_{\mu \nu \sigma \delta} \lambda^{\mu}_{(0)} \lambda^{\nu}_{(3)}
\lambda^{\sigma}_{(0)} \lambda ^{\delta}_{(3)} = [2e^{2b}b'(\eta)w,_{r} -
(1 - e^{b})^{2}w,_{rr}]/(4w).
$$
The combination of the above equation with Eqs.(3) and (6) shows that
as $r \rightarrow 0$, this component diverges like $r^{-1}$.  In other words,
the tidal forces felt by the freely-falling test particles become infinitely
large at the axis.

It might be argued that the formation of the singularity is due to the
particular solution constructed above. However, the following
considerations show that it is formed for all the radiating
CS. To see this, let us consider the spacetime described by the
metric [15]
$$ ds^2 = e^{2(K-U)}(dt^{2} - dr^{2}) - e^{2U}dz^{2} -
          e^{-2U}W^{2}d\varphi^{2},$$
where $K, U,$ and $W$ are functions of $t$ and $r$. The regularity
conditions at the axis require
\begin{equation}
|\partial\varphi|^{2} = - g_{\varphi\varphi} = e^{-2U}W^{2}
\rightarrow O(r^{2}), \;\;\;\; as \;\;\;\; r \rightarrow 0.
\end{equation}
Now, taking Eq.(4) as the source of the Einstein field equations, one
can show that the null dust has only contributions to the metric
coefficient $K(t,r)$ (for a more general discussion on this point, see
[10]). In fact, writing $K(t,r)$ as $K(t,r) =
K^{cs}(t)-b(\eta)/2$, the Einstein field equations can be divided
into two groups, one reads
\begin{equation}
\rho = b'(\eta) (W,_{r} - W,_{t})/(2 W),\;\;
\sigma = e^{2(U-K)}(W,_{tt} - W,_{rr})/W ,
\end{equation}
and the other consists of the equations for the functions $K^{cs}, W,$
and $U$, which are given exactly by those for the case where $b(\eta) =
0$ and are provided in [15]. In addition to Eq.(12), we further assume
that $\sigma$ is finite, so that when the radiation is switched off [$b(\eta)
= 0$] the spacetime is free of any singularity at the axis [16]. This
assumption together with Eq.(12) yields
$$
W(t,r) \rightarrow w_{1}r + w_{3}r^{3} + O(r^{4}),$$
\begin{equation}
e^{U(t,r)} \rightarrow w_{1} + a_{1}r + O(r^{2}), \;\;\;
as \;\;\; r \rightarrow 0,
\end{equation}
where $w_{1}, w_{3}$ and $a_{1}$ are smooth and bounded functions of $t$
only.  Note that the other Einstein equations may impose further
restrictions on these functions, however, we find that Eq.(14) is
sufficient for our present purpose. Also, since we are mainly
interested in the behavior of the spacetime near the axis, we shall
focus our attention only in the inside region of the CS, and
assume, without loss of generality, that a matching to outside is
always possible. With the above assumptions, we find that the Weyl
tensor has only three non-zero components, $\Psi_{A}\; (A = 0, 2, 4)$
in the null tetrad frame of Eq.(8), and can be
written as $\Psi_{A} = \Psi_{A}^{cs} + \Psi_{A}^{gv}$, where
$$\Psi_{A}^{cs} = - \sigma/4 + f_{A},\;\;\;\; \Psi_{2}^{gv} =
\Psi_{4}^{gv} = 0,\;\;\;\;$$
\begin{equation}
\Psi^{gv}_{0} = - e^{2(U-K)}b'(\eta)[(U,_{t} - U,_{r}) -
                (W,_{t} - W,_{r})/(2W)],
\end{equation}
and $f_{A}$ are functions of $U$ and its upto second-order
derivatives. From Eq.(14) we can see that they are finite at the axis.
In sequel, the $\Psi_{A}^{cs}$'s are finite, too.
As before, we interpret these $\Psi_{A}^{cs}$'s as representing the
gravitational field of the CS, and $\Psi^{gv}_{0}$ [which is
proportional to $b'(\eta)$] the gravitational radiation of it. In terms
of these quantities, the corresponding Kretschmann scalar is given by
\begin{equation}
{\cal R} =
16[ 3(\Psi^{cs}_{2})^{2} + \Psi^{cs}_{0} \Psi^{cs}_{4} + \sigma^{2}/6]
+ 16 \Psi^{cs}_{4}\Psi^{gv}_{0} + 4 e^{2(U-K)} \sigma \rho,
\end{equation}
where $\sigma, \rho, \Psi^{cs}_{A},$ and $\Psi^{gv}_{0}$ are given by
Eqs.(13) and (15). Similar to Eq.(11), the above equation also contains
three terms, each of which has the same physical interpretation.  In
particular, the last two terms represent the interaction of the
gravitational and matter fields of the CS with the corresponding
radiation fields emitted by it, and behave like $r^{-1}$ as
$r \rightarrow 0$ [Note that here we still have $T_{\mu
\lambda}^{cs;\lambda} = - (\rho k_{\mu}k_{\lambda})^{;\lambda}
=\frac{1}{2}\sigma b'(\eta)k_{\mu}.$]. On the
other hand, we also  have $Q_{k} \equiv - k^{\lambda};_{\lambda} =
- e^{2(U-K)}(W,_{t} + W,_{r})/W \approx -r^{-1} \rightarrow
-\infty,$ as $r \rightarrow 0$. Therefore,
it is concluded that {\em the spacetime of a radiating CS
is always singular at the axis, because of the non-linear interaction
of the gravitational and massless particle radiation fields emitted by
the CS with the corresponding ones of the CS.}

The implication of our results to CS is obvious. In
particular, it gives rise to another challenge to the structure formation
scenario from CS, unless one assumes that the gravitational
and particle radiation of ILCS is completely suppressed so that
the singularity will not be formed, or
{\em somehow} considers the singularity as physically acceptable.
 In addition, ``wiggle" CS have different properties from
the ``normal" ones, and the backreaction of the gravitational and
particle radiation of them may not necessarily produce spacetime
singularities, a problem which is now under investigation.

\vspace{.5cm}

\noindent {\bf{Acknowledgments}}

This work was partially supported by a grant from CNPq.


\end{document}